\documentclass[conference]{IEEEtran}
\IEEEoverridecommandlockouts
\usepackage{cite}
\usepackage{amsmath,amssymb,amsfonts}
\usepackage{algorithmic}
\usepackage{graphicx}
\usepackage{textcomp}
\usepackage{xcolor}
\def\BibTeX{{\rm B\kern-.05em{\sc i\kern-.025em b}\kern-.08em
    T\kern-.1667em\lower.7ex\hbox{E}\kern-.125emX}}
\begin{document}

\title{Separable-HoverNet and Instance-YOLO for Colon Nuclei Identification and Counting
\thanks{Identify applicable funding agency here. If none, delete this.}
}

\author{\IEEEauthorblockN{1\textsuperscript{st} Chunhui Lin}
\IEEEauthorblockA{\textit{Research and Development center} \\
\textit{Zhejiang Dahua Technology Co., Ltd}\\
Hangzhou, Zhejiang, China \\
lin\_chunhui1@dahuatech.com}
\and
\IEEEauthorblockN{2\textsuperscript{nd} Liukun Zhang}
\IEEEauthorblockA{\textit{Research and Development center} \\
\textit{Zhejiang Dahua Technology Co., Ltd}\\
Hangzhou, Zhejiang, China \\
zhang\_liukun@dahuatech.com}
\and
\IEEEauthorblockN{3\textsuperscript{rd} Lijian Mao}
\IEEEauthorblockA{\textit{Research and Development center} \\
\textit{Zhejiang Dahua Technology Co., Ltd}\\
Hangzhou, Zhejiang, China \\
mao\_lijian@dahuatech.com}
\and
\IEEEauthorblockN{4\textsuperscript{th} Min Wu}
\IEEEauthorblockA{\textit{Research and Development center} \\
\textit{Zhejiang Dahua Technology Co., Ltd}\\
Hangzhou, Zhengjiang, China \\
wu\_min@dahuatech.com}
\and
\IEEEauthorblockN{5\textsuperscript{th} Dong Hu}
\IEEEauthorblockA{\textit{Research and Development center} \\
\textit{Zhejiang Dahua Technology Co., Ltd}\\
Hangzhou, Zhengjiang, China \\
hu\_dong2@dahuatech.com}
}

\maketitle

\begin{abstract}
Nuclear segmentation, classification and quantification within Haematoxylin \& Eosin stained histology images enables the extraction of interpretable cell-based features that can be used in downstream explainable models in computational pathology (CPath). However, automatic recognition of different nuclei is faced with a major challenge in that there are several different types of nuclei, some of them exhibiting large intraclass variability. In this work, we propose an approach that combine Separable-HoverNet and Instance-YOLOv5 to indentify colon nuclei small and unbalanced. Our approach can achieve mPQ+ 0.389 on the Segmentation and Classification-Preliminary Test Dataset and r2 0.599 on the Cellular Composition-Preliminary Test Dataset on ISBI 2022 CoNIC Challenge\cite{b1}.
\end{abstract}

\section{Introduction}
Deep learning models have revolutionised the field of computational pathology (CPath), partly due to their ability in leveraging the huge amount of image data contained in multi-gigapixel whole-slide images (WSIs). Yet, utilizing CNNs in an end-to-end manner for slide-level prediction can lead to poor explainability, due to a high-level of model complexity with limited feature interpretability. Explainable AI in CPath may be preferable because it can ensure algorithmic fairness, identify potential bias in the training data, and ensure that the algorithms perform as expected. In order to extract meaningful human-interpretable features from the tissue, accurate localisation of clinically relevant structures is often an important initial step. For example, features indicative of nuclear morphology first require each nucleus to be segmented and can then be directly used in downstream tasks, such as predicting the cancer grade and survival analysis. Sometimes, rather than exploring features indicative of the nuclear shape, it may be of interest to accurately quantify different types of cells, which may negate the need to explicitly localise each nuclear boundary. For example, the counts of tumour cells and lymphocytes have recently been used as a powerful prognostic marker.
However, the size of cells is usually small, especially in pathological images at 20 magnification, such as 13x14. Besides, the number of different types of cells varies greatly, which may reach dozens of times. To solve these problems, we propose an approach that combine Separable-HoverNet\cite{b2} and Instance-YOLOv5 to indentify colon nuclei small and unbalanced. We first separable two HoverNets based on HRNet to identify epithelial, lymphocyte, connective and neutrophil, eosinophil, plasma respectively to solve the problem of class imbalance. The downsampling stride of the stem of HRNet\cite{b3} is modified as one to improve the recognition effect of cells with small size, and the network structure is maintained to load the pretraining weight of Imagenet. In order to impove the detection of small targets, we introduce YOLOv5 to detect cells, and realized instance segmentation through lightweight UNet\cite{b4}. Our approach can achieve mPQ+ 0.389 on the Segmentation and Classification-Preliminary Test Dataset and r2 0.5988 on the Cellular Composition-Preliminary Test Dataset.

\section{Methodology}

\subsection{Data preprocessing and augmentation}

We split the instance segmentation and semantic segmentation labels of six categories into two labels with epithelial, lymphocyte, connective and neutrophil, eosinophil, plasma to train Separable- HoverNet. Based on the rapid development of object detection, we converts instance segmentation and semantic segmentation labels into YOLO format detection labels, and cell segmentation labels for UNet. The number of less quantity categories such as neutrophils is upsampled for class balance. 

\subsection{Method description}
\subsubsection{Separable-HoverNet}
In order to solve the problem of extreme class balance, a simple method of model ensumble is adopted, which introduce two HoverNets with the same structure to indentify epithelial, lymphocyte, connective and neutrophil, eosinophil, plasma respectively. We replace the resnet50 with HRNet-w48 modify the downsampling of the stem of HRNet to improve the recognition effect of small targets. The network structure of HRNet is maintained to load the pre training weight of Imagenet. An ASPP structure is introduced after HRNet to obtain different receptive field. Three FCN-like branches connected after ASPP for horizontal and vertical distances predicting, nuclear instance segmentation and semantic segmentation.
\subsubsection{Instance-YOLO}
Compared with instance segmentation networks such as Mask RCNN\cite{b5} and SOLO\cite{b6}, we found YOLOv5 has better performance on small cells detection. In order to realize the segmentation task of the instance, we combined the cell segmentation of UNet and the cell detection of YOLOv5. It is worth mentioning that we have modified the down sampling structure of UNet because the input resolution of 16x16 has more accurate boundaries. Equalized Focal Loss\cite{b7} is used to alleviate the imbalance of various inflammatory cells by rebalances the loss contribution of positive and negative samples of different categories independently according to their imbalance degrees. A customized-mosaic enhancement method was used to ensure that the three grids of mosaic have neutrophils, eosinophils and plasma respectively to achieve a certain degree of class balance.
\subsection{Post Processing}
Two HoverNet results are firstly fused through simple result superposition. The image post-processing process aims to refine the result of bounding boxes and UNet cell segmentation and convert it to instance and segmentation map. After getting the results of HoverNet and YOLOv5, these results are fused by overlapping the results of YOLOv5 on that of HoverNet in the event of HoverNet misses this cell.

\section{Experiment}
\subsection{Dataset}
Our algorithm is evaluated on the ISBI 2022 CoNic challenge. The Lizard\cite{b8} data from the following dataset sources: DigestPath, CRAG, GlaS, CoNSeP and PanNuke, consisting of a total of 431,913 labelled nuclei. Of these nuclei, 210,372 are epithelial, 92,238 are lymphocytes, 24,861 are plasma cells, 4,116 are neutrophils and 2,979 are eosinophils. 
\subsection{Experiment Setup}
A sliding window scheme with overlap is used to crop each WSI into small patches of size 256x256 pixels. Standard real-time data augmentation methods such as horizontal flipping, vertical flipping, random rescaling, random cropping, and random rotation are performed to make the model invariant to geometric perturbations. The Adam optimizer is used as the optimization method for model training. The initial learning rate is set to 0.001, and reduced by a factor of 10 at the 25th epoch, with a total of 50 training epochs in HoverNet. We trained YOLOv5 from scratch for 150 epochs and finetune 100epochs by SGD optimizer. The network of HoverNet an UNet are trained by minimizing a total loss function composed of a Cross entropy Loss and a Dice loss. The network of YOLOv5 are trained by minimizing a total loss function composed of a Cross entropy Loss and a EFL loss for classification and CIoU loss for bounding boxes regression. Test time augment is used during inference in both YOLO and UNet.
\section{Results}
Performance of two models are reported in Table 1. The first two rows are results of HoverNet and our model respectively. Comparing the second row, our model achieves stronger performance on both Segmentation and Classification and Cellular Composition Tasks. It outperforms HoverNet 0.097 and 1.027 of mPQ+ and r2, which indicates the ensemble of Separable-HoverNet and Instance-YOLOv5 enhances the small cells Identification.

\begin{table}[htbp]
\caption{Results on Preliminary Test}
\begin{center}
\begin{tabular}{|c|c|c|}
\hline
 Model & \textbf{mPQ+} & \textbf{r2} \\
  & Segmentation and Classification& Cellular Composition \\
\hline
 HoverNet & 0.296 & -0.428 \\
\hline
 ours & 0.389 & 0.599 \\
\hline
\end{tabular}
\label{tab1}
\end{center}
\end{table}

\end{document}